\documentclass{article}
\usepackage{authblk}
\usepackage[a4paper, left=1.3in, right=1.3in,]{geometry}

\title{A Differentiable Surrogate Model for the Generation of Radio
Pulses from In-Ice Neutrino Interactions}

\date{}

\author[1]{Philipp Pilar\thanks{Corresponding author: philipp.pilar@it.uu.se}}%
\author[2]{Martin Ravn}
\author[2,3]{Christian Glaser}
\author[1]{Niklas Wahlström}

\affil[1]{Department of Information Technology,
Uppsala University, Uppsala, Sweden}
\affil[2]{Department of Physics and Astronomy, Uppsala University, Uppsala, Sweden}
\affil[3]{Department of Physics, TU Dortmund University, Dortmund, Germany}

\usepackage{amsmath}

\usepackage{amsfonts}       
\usepackage{amssymb}
\usepackage{mathtools}
\usepackage{wrapfig}
\usepackage{booktabs}
\usepackage{caption}
\usepackage{hyperref}
\usepackage{color}
\usepackage{pythonhighlight}
\usepackage{makecell}
\usepackage[round]{natbib}
\usepackage{wrapfig}

\usepackage{mynotation} 

\begin{document}

\maketitle

\begin{abstract}
The planned IceCube-Gen2 radio neutrino detector at the South Pole will enhance the detection of cosmic ultra-high-energy neutrinos.
It is crucial to utilize the available time until construction to optimize the detector design.
A fully differentiable pipeline, from signal generation to detector response, would allow for the application of gradient descent techniques to explore the parameter space of the detector.
In our work, we focus on the aspect of signal generation, and propose a modularized deep learning architecture to generate radio signals from in-ice neutrino interactions conditioned on the shower energy and viewing angle.
The model is capable of generating differentiable signals with amplitudes spanning multiple orders of magnitude, as well as consistently producing signals corresponding to the same underlying event for different viewing angles.
The modularized approach ensures physical consistency of the samples and leads to advantageous computational properties when using the model as part of a bigger optimization pipeline.
\end{abstract}

\section{Introduction}

The detection of high-energy neutrinos opens up a unique way of perceiving the depths of the cosmos \citep{meszaros2019multi}.
Due to their close-to-vanishing interaction with matter, they can traverse vast distances essentially undisturbed.
Hence, as messengers, they carry the potential to unambiguously identify the location of cosmological events.
However, their weak interaction also makes them extraordinarily difficult to detect, and the instrumentation of huge volumes is required to obtain acceptable detection rates.
A cost-efficient way to achieve this is to instrument vast ice shelves in Greenland or Antarctica with an array of compact radio detector stations \citep{Barwick:2022vqt}.
The big volume of detector material, combined with the high transparency of ice, allows for the occasional detection of high-energy neutrinos.

We focus our work on the radio detector of the IceCube-Gen2 project \citep{aartsen2021icecube, TDR}, which is currently in development for construction at the South Pole and whose design can still be optimized.
The potential improvement is large. In \cite{Glaser:2023udy}, under optimistic assumptions, it was estimated that an improved station layout, together with additional trigger improvements, can lead to an up to three-times better detector at negligible additional costs. 
To maximize the efficacy of the detector design, the precise locations and orientations of the radio antennas need to be optimized.
The differentiable programming paradigm \citep{baydin2018automatic, blondel2024elements} constitutes an intriguing option to address this problem.
Constructing a fully differentiable pipeline, from neutrino event to detector response, would allow for the use of gradient descent methods for optimization.
In this work, we consider the aspect of radio pulse generation and design a differentiable surrogate model to replace (parts of) the existing Monte Carlo simulations.

\begin{figure}
    \centering
    \includegraphics[width=\linewidth]{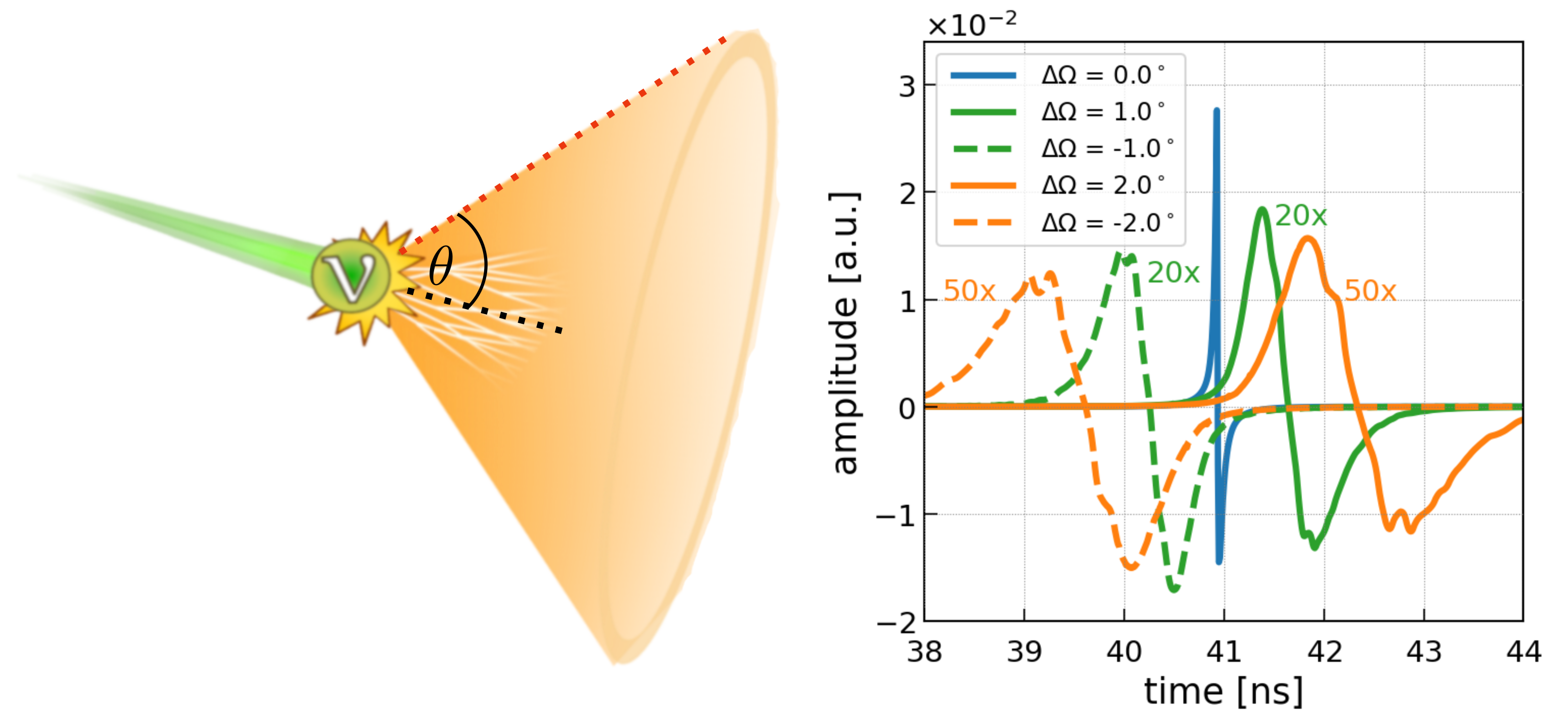}
    \caption{
    Figures reproduced from \citet{Barwick:2022vqt} with permission\protect\footnotemark.
    \textbf{Left:}
    A neutrino interaction in ice.
    The collision of a neutrino with an atom in the ice gives rise to a shower of secondary particles.
    These particles emit Askaryan radiation, which is strongest on the Cherenkov cone.
    In this plot, the viewing angle $\theta=\theta_C$, where $\theta_C$ denotes the Cherenkov angle in ice, and where the dotted red line indicates the direction towards the observer.
    \textbf{Right:}
    The same signal is depicted at different viewing angles, where $\Delta \Omega = \theta - \theta_C$.
    It can be observed that the signals with $\pm \Delta \Omega$ are approximately antisymmetric copies of each other.
    }
    \label{fig:askaryan}
\end{figure}

\footnotetext{Radio Detection of High Energy Neutrinos in Ice, Steven Barwick and Christian Glaser, Edited by Floyd W. Stecker, The Encyclopedia of Cosmology, Vol 2 Chapter 6, Copyright @ 2023 World Scientific Publishing.}

The radio pulses originate from neutrino interactions in ice \citep{Barwick:2022vqt}.
A cascade of secondary particles forms a particle shower with time-varying excess negative charge, which gives rise to Askaryan radiation (see Section \ref{sec:Askaryan} for details).
A schematic of the process is depicted in the left part of Figure \ref{fig:askaryan}.
The radiation is strongest when observed on the Cherenkov cone with viewing angle $\theta=\theta_C$, which is 56$^\circ$ in deep ice.
In the right plot of Figure \ref{fig:askaryan}, signals from the same event at different viewing angles are depicted.
The signals exhibit an approximate antisymmetry around $\theta_C$, and the amplitude decreases quickly when $\theta$ diverges from it.
The neutrino energy $E$ determines the underlying signal shapes that are possible.
While the distance between the event and the observer also affects the signal, we do not consider this aspect in this work and instead use a constant distance of $1\,\text{km}$.

\begin{figure}[!t]
    \centering
    \includegraphics[width=1\linewidth]{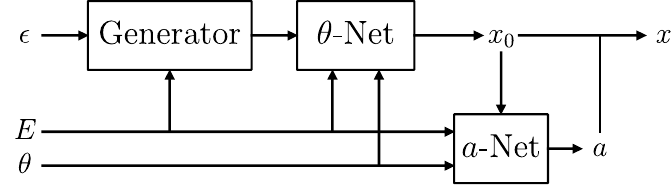}
    \caption{
    The modularized model architecture.
    The generator generates normalized samples at a fixed angle and for a given shower energy $E$.
    The $\theta$-Net then transforms the signal to the desired viewing angle $\theta$.
    Finally, the $a$-Net predicts the signal amplitude $a$ and combines it with the normalized signal $x_0$ to obtain the signal $x$.
    }
    \label{fig:modularized_model}
\end{figure}

Additional requirements are placed on the generative surrogate model due to its usage as a component in the detector optimization pipeline.
It needs to generate signals together with their respective amplitudes, which range over many orders of magnitude.
Furthermore, it needs to be able to generate multiple correlated signals, corresponding to the same neutrino interaction as observed by antennas at different positions.
To fulfill these requirements, we develop a modularized model architecture, which is illustrated in Figure \ref{fig:modularized_model}.

A generative model (the generator) producing normalized signals lies at the core of this architecture. 
The type of model used for the generator can be freely chosen, as long as realistic samples are generated.
Even non-differentiable simulations, such as the Monte Carlo simulation itself, are possible;
it is one of the advantages of the modularized architecture to enable (approximate) differentiability for the event parameters also in such cases.
In most of our experiments, we utilized the MC simulation as the generator.
To illustrate certain aspects of the model architecture, we also employed a diffusion model \citep{ho2020denoising} as an alternative.
The normalized signals are subsequently transformed to the different viewing angles of interest and combined with their final amplitudes.
Two additional networks are employed for these steps.

\section{Background}

In this section, we give background information on the dataset and the physical processes involved in giving rise to the radio signals under consideration.

\subsection{Askaryan radiation} \label{sec:Askaryan}
Askaryan radiation was first described in \citet{askaryan1965coherent}.
Neutrinos can occasionally interact with atoms in the ice and cause a cascade of secondary particles, called a particle shower.
As the particle shower develops, electrons outnumber positrons, and a negative charge excess builds up along its trajectory, resulting in the generation of radio emission. 
Since the particle shower is traveling faster than the speed of light in ice, the emitted radiation exhibits Cherenkov-like characteristics and is strongest on the Cherenkov cone, with viewing angle $\theta = \theta_C$, where $\theta_C=55.82^\circ$ is the Cherenkov angle in deep ice.
See Figure \ref{fig:askaryan} for an illustration of this process.

While the development of the particle shower is stochastic, its main characteristics depend on the neutrino energy and the type of interaction.
The higher the neutrino energy, the more secondary particles can be created, and the larger the negative charge excess becomes.
In hadronic showers, the first few interactions are hadronic, before electrons and positrons are created via pair production.
Here, the charge-excess profile exhibits a single peak and does not differ significantly between different energies, apart from the amplitude.
In electromagnetic showers, the neutrino interaction immediately gives rise to an electron, and subsequently electron-positron pairs.
Due to the Landau-Pomerachuk-Migdal (LPM) effect \citep{landau_pomeranchuk_1953}, electrons can travel further between interactions at higher energies.
Multiple spatially separated showers can emerge, which result in charge-excess profiles with multiple peaks, and, in turn, more complicated radio signals.
A more detailed description of the processes giving rise to Askaryan radiation can be found in \citet{Barwick:2022vqt}.

\subsection{Dataset} \label{sec:data}

The data under consideration in this project consists of 1D waveforms generated with the NuRadioMC library \citep{glaser2020nuradiomc} using the ARZ2020 model \citep{alvarez2011practical}.
We consider electron-neutrino charged current interactions.
Energies lie in the range $\log_{10} E \text{[eV]} \in [15,19]$, and viewing angles in $\theta \in [35.82^{\circ}, 75.57^{\circ}]$.
Due to the aforementioned stochasticity in the shower development, different shower profiles are possible at the same energy.
In each $0.1 \log_{10} E \text{[eV]}$ bin, the NuRadioMC library provides 10 different shower profiles.
The viewing angle $\theta$ can be freely chosen in the stated range (for values outside that range, the signal amplitude becomes vanishingly small) and we generated data on a grid covering the range stated above with 160 values.
Hence, our dataset consists of $65\,600$ samples, in total, but there are only 410 truly independent signals.
Exemplary samples are depicted, e.g., in Figures \ref{fig:askaryan} and \ref{fig:theta_relation}.

For each of these samples, the following data is available:
the waveform $x$ on a time grid of size $1024$ corresponding to a sampling rate of $5 \times 10^{-10} [s]$, together with the event energy $E$, and the viewing angle $\theta$. 
We denote the amplitude as $a$, and the normalized sample with amplitude unity as $x_0$.
When used as inputs to the neural networks, the values of $E$ and $\theta$ are normalized to lie in the range $[0,1]$ via $E \rightarrow \frac{\log_{10} E \text{[eV]} - 15}{4}$ and $\theta \rightarrow \frac{\theta [^\circ] - 35.82}{39.75}$.

When training our models, we split the data into train and test data.
To make the test data truly independent, when a signal was added to the test data, all the corresponding signals with the same energy and shower profile at different angles were added to the test data as well.
In total, $90\%$ of the data was used for training, and $10\%$ was held out for testing.
This proportion holds at all energies.

\section[Model architecture]{Model architecture
\protect\footnote{The code for the project is available on GitHub: \href{https://github.com/ppilar/IC_Surrogate}{https://github.com/ppilar/IC\_Surrogate}.} \label{sec:model}} \label{sec:architecture}

A schematic of the architecture of our generative model is given in Figure \ref{fig:modularized_model}.
Machine learning models generally perform best when inputs and outputs lie in a narrow range of values \citep{huang2023normalization}.
For the $E$ and $\theta$ values, this can be straightforwardly achieved by normalizing the data, as described in the previous section.
The large range in which the amplitudes of the radio pulses lie (compare Section \ref{sec:data}), however,  poses a challenge.
Due to the zero-crossings of the samples, taking the logarithm as we did for $E$ to reduce the range of the numerical values before normalizing is not possible.
For this reason, we have introduced the amplitude net ($a$-Net) to predict the signal amplitudes based on their shape, as well as the viewing angle $\theta$ and the neutrino energy $E$.
This allows us to use normalized data with amplitude unity for each sample when training the different model components.

Another requirement for the model is the ability to generate the same signal when viewed from different angles.
This is relevant when considering the case where multiple antennas located at different positions detect the same neutrino interaction.
While the generative model may automatically learn the correct relationship from the data, this is not certain, and we want to add a stronger inductive bias to ensure correct model behavior.
We achieve this by further splitting the generative model into the generator, which now generates normalized signals at a fixed reference angle $\theta_r$, and the $\theta$-Net, which transforms the thus-generated signals to the desired angle.

Apart from ensuring more accurate sample generation, the modularized architecture has additional advantages when viewed from the perspective of a differentiable optimization pipeline.
This becomes apparent when considering the computational path that the inputs take in Figure \ref{fig:modularized_model}.
The viewing angle $\theta$ does not pass through the generator, and the computational graph only needs to be created for the $\theta$-Net and the $a$-Net when computing gradients.
While the generative model still takes the energy as input, the strongest contributions to the energy derivatives are expected to originate from the amplitude.
Hence, approximate gradients can be obtained efficiently also for $E$ when neglecting the contribution from the generator.
This approximation enables the use of expensive generative models like diffusion models with many denoising steps.
It also enables the use of non-differentiable models for the generator, such as Monte Carlo simulations.

\subsection{A diffusion model as generator}

\begin{figure}
    \centering
    \includegraphics[width=1\linewidth]{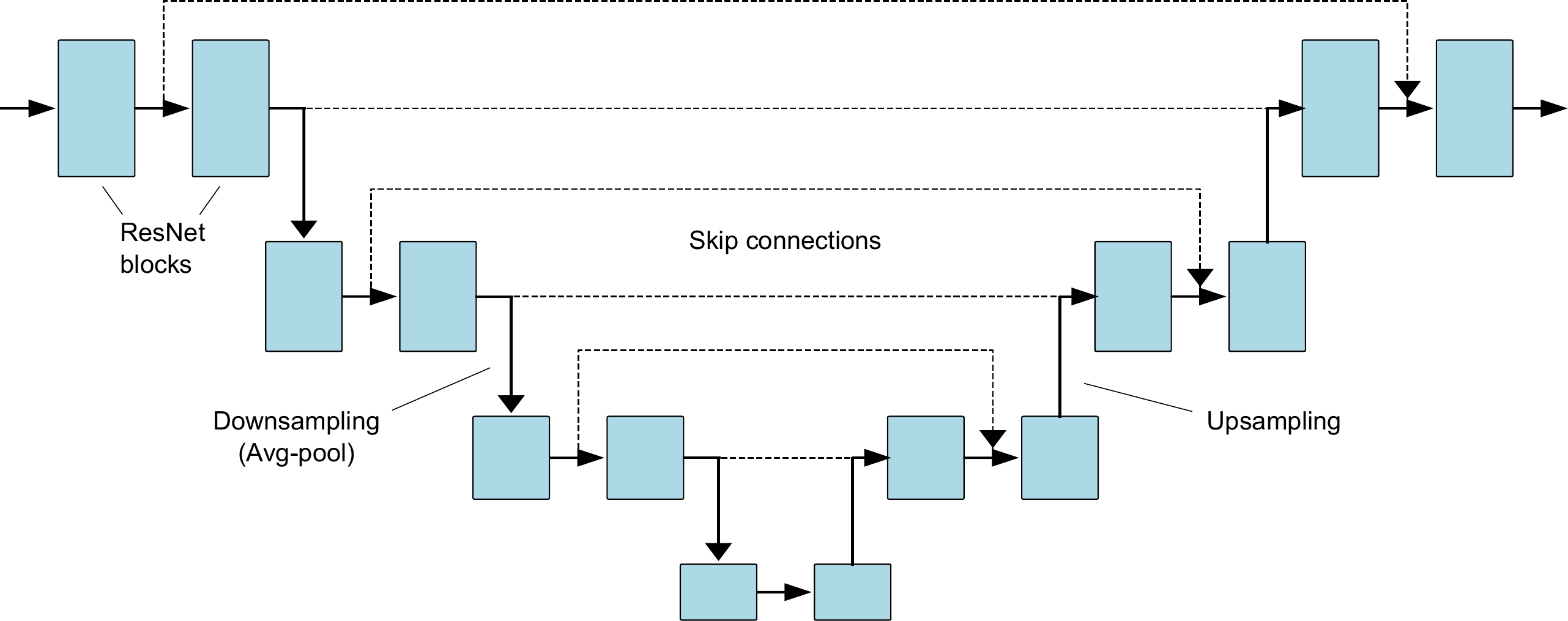}
    \caption{
    The U-Net architecture \citep{ronneberger2015u} is often employed for transformations with the same input- and output size.
    It consists of multiple levels that allow for the extraction of features at different scales.
    Each level contains multiple ResNet blocks (the blue boxes). 
    When moving up or down between levels, the outputs are down- or upsampled, respectively.
    Skip connections connect the different levels on both sides, from the encoder (left) to the decoder (right), enabling more stable learning.
    We employ U-Nets for the fine-tuning in the $\theta$-Net and the denoiser in the diffusion model.
    }
    \label{fig:UNet}
\end{figure}

While existing MC simulations can be used for the generator, we also evaluate the option of using a generative machine learning model in their stead.
Denosing diffusion probabilistic models (DDPMs, \citet{ho2020denoising}) generate samples from Gaussian noise $\epsilon$ by removing the noise stepwise, as outlined in Appendix \ref{app:ddpm}.
The denoiser performing this task is its centerpiece, and the U-Net architecture \citep{ronneberger2015u} depicted in Figure \ref{fig:UNet} is usually employed for it.
Due to their constituent convolutional layers, U-Nets are well-suited for spatiotemporal data.
The different levels of the U-Net enable the extraction of features at different scales, with skip connections helping to retain information from extracted features at higher levels of the U-Net.
In our implementation, we also chose the U-Net architecture and based it on the variant that comes with the python package by  \citet{lucidrains2020denoising}.

Each of the U-Net layers contains multiple ResNet blocks.
An average-pooling layer is then utilized to downsample to the layer below the current one;
in the upward pass, an upsampling layer is employed instead.
Skip connections connect to the opposite side of the U-Net, both from the middle and the end of each U-Net level.
The ResNet blocks contain multiple convolutional layers, supplemented with skip connections, as well as a rescaling operation.
Scale and shift for the rescaling operation are calculated via a fully-connected layer from the embedding of the diffusion step, which is used to condition on the current step $k$ in the denoising process.

We have chosen 6 levels for the U-Net, with [8, 16, 32, 64, 128, 256] channels in the convolutional layers, respectively.
The Adam optimizer with learning rate $8 \times 10^{-5}$ was employed, and the network was trained until convergence for $5 \times 10^{5}$ iterations with batch size 128.
For the diffusion and denoising process, 500 steps were chosen.
When sampling signals, the effective number of steps taken can be reduced \citep{salimans2022progressive}, and we use 50 steps when sampling from the trained DDPM.

In our model, we have modified the step embedding to also condition on the neutrino energy $E$ and, optionally, the viewing angle $\theta$.
This is done by first using three different nets to find separate embeddings for each of these quantities, followed by another network to combine the embeddings.
The individual embedding networks each consist of a positional encoding layer of dimension $16$ followed by two linear layers with outputs of size $32$. 
The GELU activation function is applied between the linear layers.
Then, the embeddings are combined via another neural network consisting of two linear layers with the GELU activation function in between.

\subsection{The $a$-Net}

\begin{figure}
    \centering
    \includegraphics[width=0.8\linewidth]{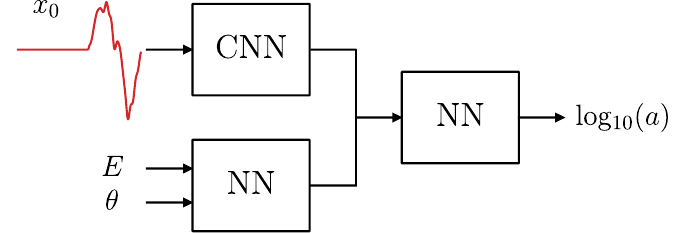}
    \caption{
    The architecture of the amplitude net ($a$-Net) is depicted.
    Features are first extracted from normalized signals and then combined with an embedding of $E$ and $\theta$.
    Then, these features are combined in a third network to yield log-amplitude predictions.}
    \label{fig:aNet}
\end{figure}

Figure \ref{fig:aNet} depicts the architecture of the $a$-Net, which consists of three separate subcomponents.
First, the normalized waveforms $x_0$ are fed through a convolutional network to extract features.
These features are then concatenated with a learned embedding of the neutrino energy $E$ and the viewing angle $\theta$.
Together, they are subsequently passed through another fully-connected neural network that predicts the logarithm of the amplitude.

The convolutional network consists of 3 layers with 32, 64, and 128 channels, respectively.
The corresponding kernels have size 3, stride 1 and zero-padding enabled.
After each convolutional layer, an average-pooling layer with kernel size 2 and stride 2 is applied, effectively halving the size of the layer output.
After the final convolutional layer, the output is flattened into an array of size $16\,384$.
The embedding network contains 3 layers of width 40.
The third network contains 3 layers of width 256, 128, and 1, respectively.
The ReLU activation function is applied after each layer, apart from average-pooling layers.

The mean squared error (MSE) loss,
\begin{equation}
    \mathcal{L}_{a} = \frac{1}{N_{\rm bs}} \sum_{i=1}^{N_{\rm bs}} (\log_{10}(\hat a_i) - \log_{10}(a_i))^2,
\end{equation}
was used for training the $a$-Net.
The Adam optimizer was used with learning rate $0.0001$ and batch size $N_{\rm bs} = 128$.
The training lasted $150\,000$ iterations, until convergence.

\subsection{The $\theta$-Net}

\begin{figure}
    \centering
    \includegraphics[width=1\linewidth]{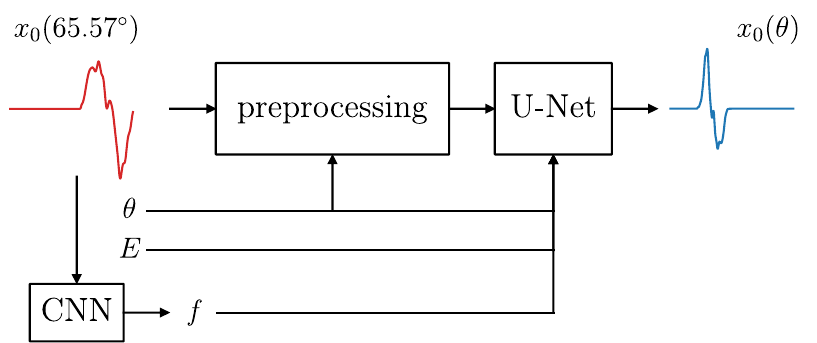}
    \caption{
    The architecture of the $\theta$-Net.
    First, the normalized signals at the fixed reference angle $\theta_r=65.57^{\circ}$ are preprocessed via Equation~\eqref{eq:preprocessing}.
    A CNN is employed to extract additional features from the reference signal.
    Then, the preprocessed signals are finetuned using a U-Net, to yield normalized signals $x_0(\theta)$ at the desired angle $\theta$.
    }
    \label{fig:tNet}
\end{figure}

The $\theta$-Net transforms reference signals at the fixed angle $\theta_r = 65.57^{\circ}$ to the desired viewing angle $\theta$.
The precise choice for the angle $\theta_r$ is essentially arbitrary, the main condition being that all relevant details of the signals must be resolved well at this angle (compare, e.g., Figure~\ref{fig:theta_relation}).
The architecture of the $\theta$-Net is depicted in Figure \ref{fig:tNet}, and consists of two main blocks, one for preprocessing and one for fine-tuning the signals.
Furthermore, a small convolutional network is employed to extract additional features from the reference signals.
In Appendix \ref{app:tnet_ablation}, an ablation study is conducted to show the impact of these components.

The main purpose of the preprocessing block is to simplify the task of the subsequent U-Net.
The normalized reference signals are transformed by using a rough geometric relationship obeyed by the signals.
As can be seen in Figure \ref{fig:theta_relation}, when starting at a large viewing angle, the signals first get squeezed when approaching the Cherenkov angle $\theta_C=55.82^{\circ}$, then flipped upside down after crossing it, and finally unsqueezed again when moving on to lower values $\theta$.
The following expression captures this behavior well:
\begin{equation} \label{eq:preprocessing}
    \tilde x_0(t, \theta) = 
    \begin{cases}
x_0 \left(f(t-t_0) + t_0, \theta_r \right) & \text{if } \theta \geq \theta_C, \\
1 - x_0 \left(0.75f(t_0-t) + t_0, \theta_r \right)  & \text{if } \theta < \theta_C, \\
\end{cases}
\end{equation}
where $t_0=256 \text{ns}$ is the middle point of the chosen time grid (approximately, where signals
\begin{wrapfigure}{r}{0.42\textwidth}
    \centering
    \vspace{0cm}
    \includegraphics[width=\linewidth]{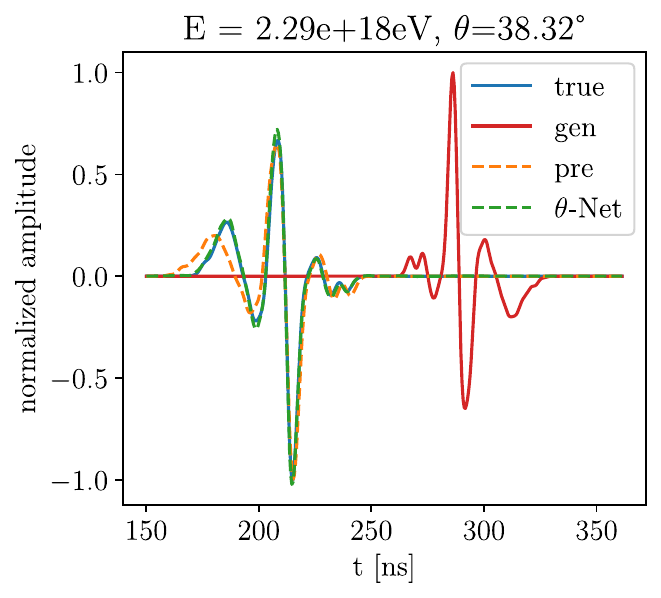}
    \caption{
    Illustration of the modularized sample generation via the $\theta$-Net.
    The generator (gen) generates the reference signal at a fixed angle.
    Preprocessing (pre) based on an approximate geometric relationship transforms the signal to a shape close to the true one.
    The U-Net then finetunes the signal shape to yield the final $\theta$-Net prediction ($\theta$-Net).
    Typically, results very close to the true curve (true) are achieved.
    }
    \label{fig:tNet_sample}
    \vspace{0cm}
\end{wrapfigure}
 with $\theta=\theta_C$ have their maximum), and where $f = \text{max}(\lvert\frac{\theta - \theta_C}{\theta_r - \theta_C}\rvert, 0.1)$ is an empirically determined squeezing factor.
The minimum value of $0.1$ prevents signals from being squeezed to almost zero width close to $\theta_C$, thereby losing the information inherent in the sample.
The additional factor $0.75$ for $\theta < \theta_C$ reflects the fact that the signals appear to be squeezed more strongly in this regime.
Both factors were determined empirically.

The convolutional network is employed to extract features from the reference signals, which may get lost when transforming the signal.
These features are then used together with $E$ and $\theta$ to condition the U-Net.
Since the reference signals are zero on the left half of the time axis, only the right half is passed as input to the CNN.
The CNN consists of three convolutional layers, followed by a flattening operation and 3 fully-connected layers.
The convolutional layers have [16,32,64] channels, respectively, a kernel size of 4, and a stride of 4.
After the flattening operation, the outputs contain 512 features.
The linear layers then map to [100, 50, 32] hidden units, respectively.
The ReLU activation function is applied after each of these layers, except the last one.

Subsequently, the U-Net architecture depicted in Figure \ref{fig:UNet} is employed for fine-tuning the signals.
The U-Net is very similar to the one used in the diffusion model, which is a slightly modified version of the one from \citet{lucidrains2020denoising}.
For the $\theta$-Net, we chose to use 5 U-Net levels with [8, 16, 32, 64, 128] channels, respectively.
The ResNet layers are conditioned on the parameters $E$ and $\theta$ in the same way as before in the diffusion model.
In contrast to the diffusion model, the U-Net is only applied once here.
Since there is no denoising step involved in this case, a constant value is used instead.
Figure \ref{fig:tNet_sample} illustrates the signal generation via the $\theta$-Net.

The loss function
\begin{equation}
    \mathcal{L}_{\theta} = \frac{1}{N_{\rm bs}} \sum_{i=1}^{N_{\rm bs}} \int (\hat x_{0i}(t, \theta_i) - x_{0i}(t, \theta_i))^2 \mathrm{d}t
\end{equation}
was employed for the $\theta$-Net.
The Adam optimizer was used with learning rate $0.0003$ and batch size 128.
The model was trained for $300\,000$ iterations until convergence.

\section{Results}

In this section, we investigate the performance of the individual model components.
We verify that the modularized architecture accurately reproduces the physical dependence of the signal shape on the viewing angle, together with realistic amplitudes.
In a simple optimization experiment, we demonstrate the usability of the model architecture in future optimization pipelines.
Additional results for the DDPM as generator are given in Appendix \ref{app:ddpm}.

\subsection{Amplitude predictions}

\begin{figure}[t]
    \centering
    \includegraphics[width=0.9\linewidth]{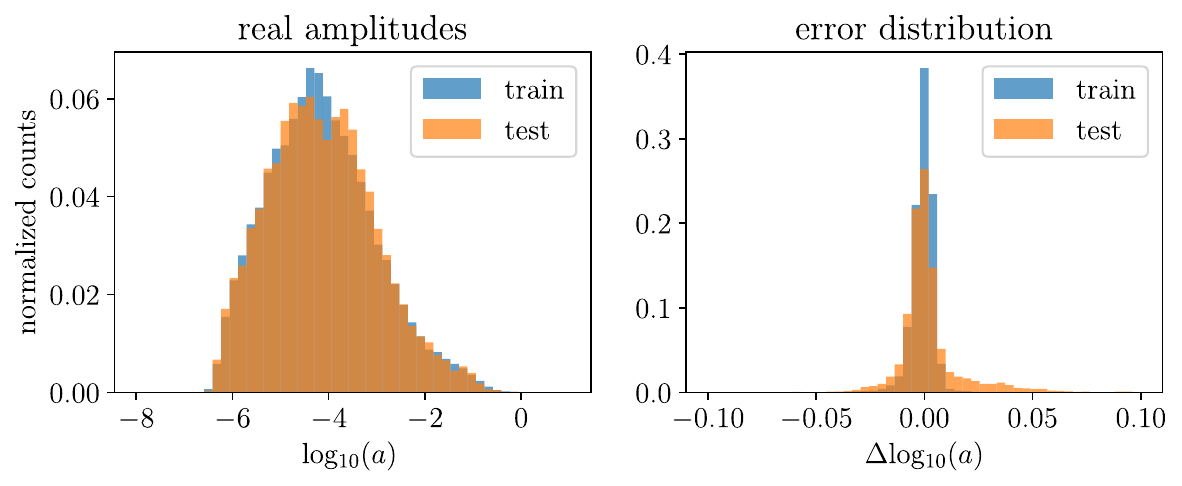}
    \caption{
    Performance of the $a$-Net.
    \textbf{Left:}
    The distribution of the signal amplitudes for training and test data.
    \textbf{Right:} 
    The distribution of the errors in the model predictions of the log amplitudes for training and test data.}
    \label{fig:aNet_results}
\end{figure}

In Figure \ref{fig:aNet_results}, the distributions of amplitudes in the training and test data are depicted, together with the error distributions of the $a$-Net predictions.
To evaluate the accuracy of the $a$-Net, we chose the distance of true and predicted values of $a$ in log-space, i.e., $\Delta \log_{10}(a) = \log_{10}(a) - \log_{10}(\hat a)$.
The 0.95-quantile of the absolute error in log space, $|\Delta \log_{10}(a)|$, lies at $0.042$ for the test data.
This means that in 95\% of the cases, the error in the $a$-Net predictions of $a$ is below roughly $10\%$ of the corresponding signal amplitudes.
The full evaluation metrics are shown in Table~\ref{tab:a_eval}.

\begin{table}[t]
\caption{
Evaluation metrics for the $a$-Net.
For $\Delta \log_{10} (a)$, the mean plus-or-minus one standard deviation is given.
For $|\Delta \log_{10} (a)|$, the 0.95-quantile is given.
}
\label{tab:a_eval}
\centering
\resizebox{0.4\textwidth}{!}{\begin{tabular}{lcc}
\toprule
\textbf{$a$-Net} & $\Delta \log_{10}(a)$ & $|\Delta \log_{10}(a)|$  \\
\midrule
train & 0.002$\pm$0.004  & 0.008 \\
test & 0.007$\pm$0.017 & 0.042 \\
\bottomrule
\end{tabular}}
\end{table}

\subsection{$\theta$-dependence} \label{sec:tnet_eval}

\begin{figure}[t!]
    \centering
    \includegraphics[width=1\linewidth]{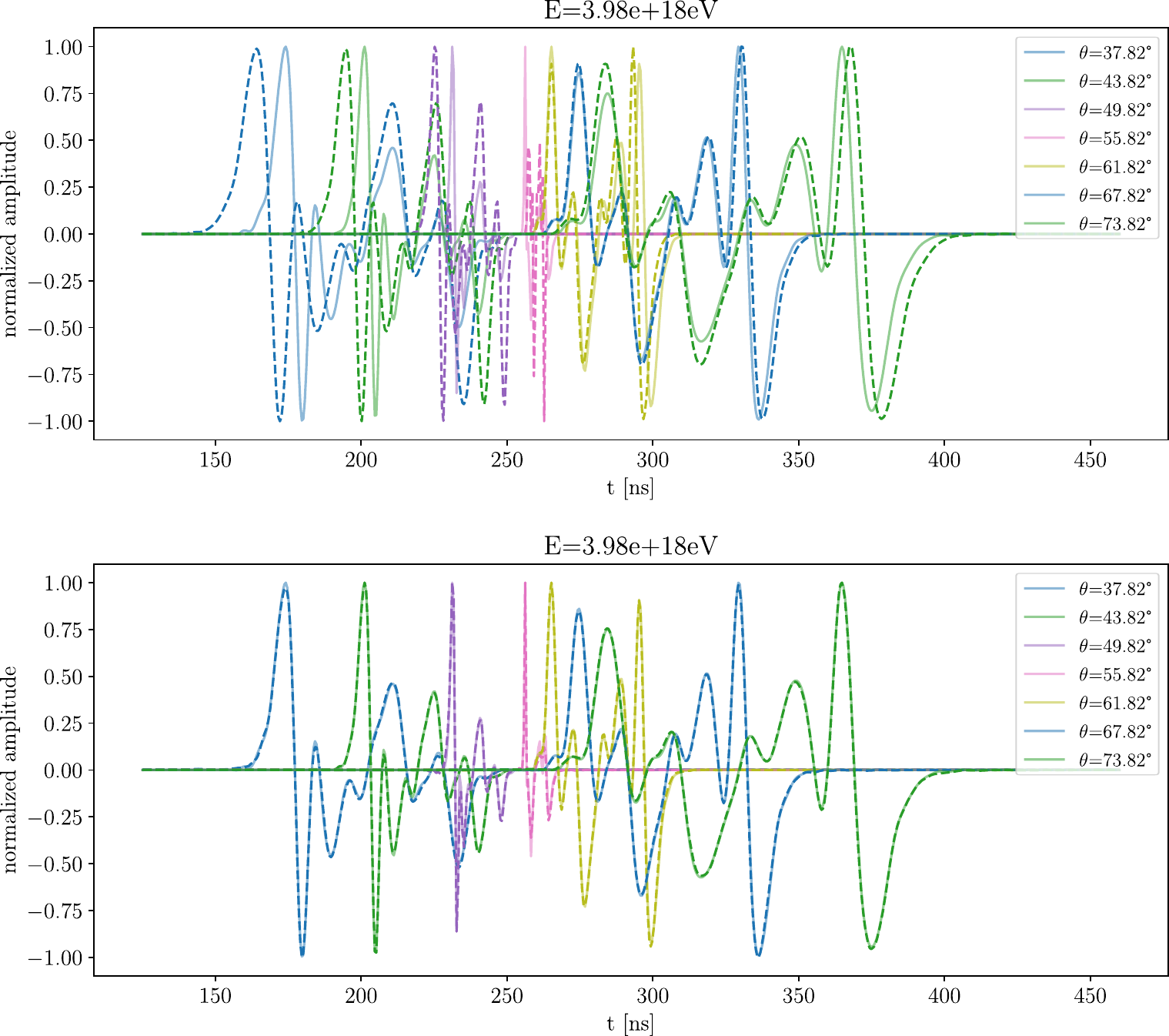}
    \caption{
    Illustrating the $\theta$-dependence.
    \textbf{Top:}
    The $\theta$-dependence after the preprocessing step.
    The solid lines are the true signals, and the dashed lines are the signals after preprocessing.    
    \textbf{Bottom:}
    The $\theta$-dependence when using the full $\theta$-Net, with both preprocessing and fine-tuning.
    The solid lines are the true signals, and the dashed lines are the signals after applying the $\theta$-Net.
    }
    \label{fig:theta_relation}
\end{figure}

In Figure \ref{fig:theta_relation}, the same signal from the test data when viewed at different angles is depicted, once after the preprocessing step and once when using the full $\theta$-Net.
It can be observed that the preprocessing succeeds at getting the signal reasonably close to the final signal, but is not yet accurate enough without the fine-tuning via the U-Net.

\begin{figure}[t!]
    \centering
    \includegraphics[width=1\linewidth]{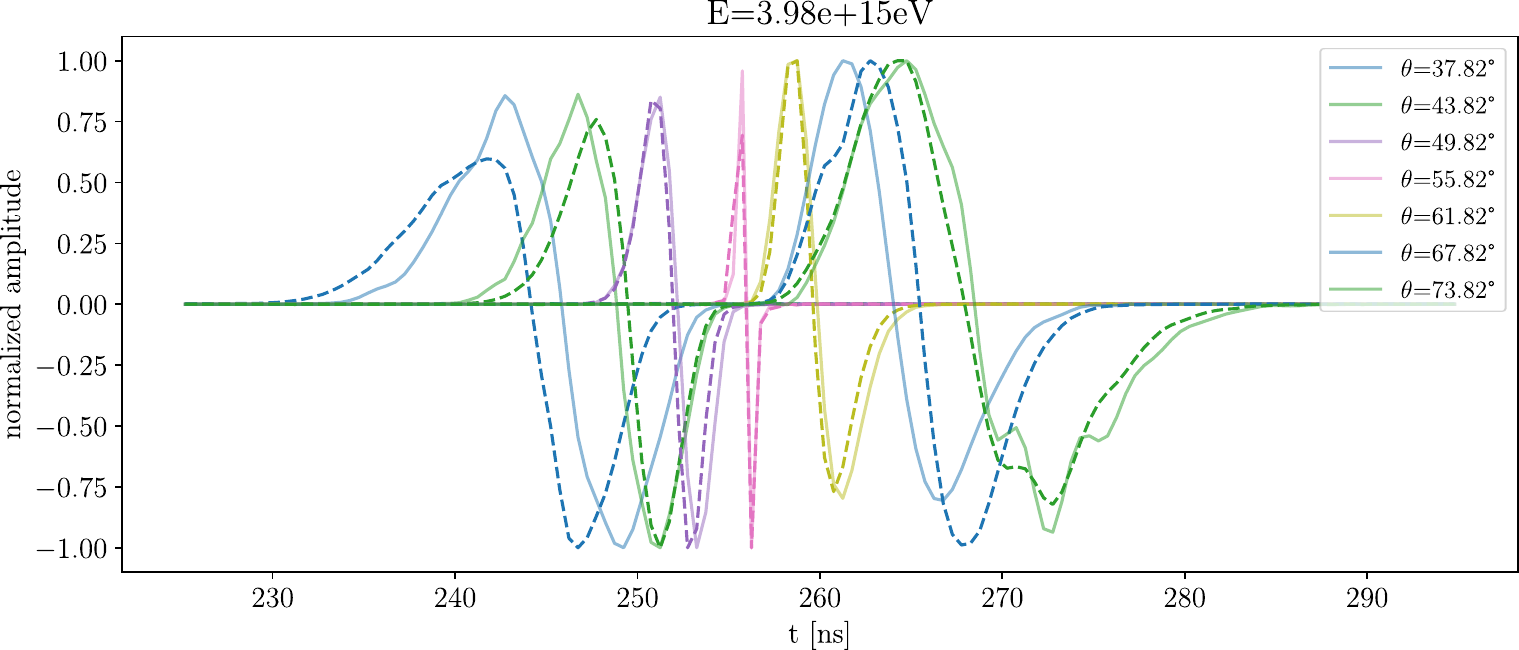}
    \caption{
    Illustrating the $\theta$-dependence when using a DDPM as the generator without a separate $\theta$-Net.
    The solid lines are the true signals, and the dashed lines are the signals generated by the DDPM.
    }
    \label{fig:DDPM_theta_relation}
\end{figure}

In Figure \ref{fig:DDPM_theta_relation}, an equivalent plot is given for the case where a DDPM is used as the generator, and where no separate $\theta$-Net is employed.
It is apparent that the DDPM alone does not automatically learn the correct $\theta$-dependence.
It roughly reproduces the locations of the peaks, but fails to accurately reproduce the finer features of the signals.
This shows that the $\theta$-Net is also required to ensure sample consistency when using a generator other than the MC simulation, in addition to bestowing properties advantageous for the model differentiability.

We quantify the accuracy of the $\theta$-Net via the normalized absolute difference between the true and predicted curve, integrated over the time domain,
$\Delta x_0=\frac{\int_{-\infty}^{\infty} |x_0(t, \theta) - \hat x_0(t, \theta)|\mathrm{d}t}{\int_{-\infty}^{\infty} |x_0(t, \theta)|\mathrm{d}t}$,
as well as the difference along the $t$-axis between the positions of the true and predicted maximum of the signal,
$\Delta \max x_0 = \underset{t}{\text{argmax}} \,\, \hat x_0(t, \theta) - \underset{t}{\text{argmax}} \,\, x_0(t, \theta)$ .
We also consider the analogous metrics in Fourier space, where $\tilde x_0$ denotes the Fourier transform of $x_0$:
$\Delta \tilde x_0 =\frac{\int_{-\infty}^{\infty} |\tilde x_0(f, \theta) - \hat{\tilde x}_0(f, \theta)|\mathrm{d}f}{\int_{-\infty}^{\infty} |\tilde x_0(f, \theta)|\mathrm{d}f}$, and 
$\Delta \max \tilde x_0 = \underset{f}{\text{argmax}} \,\, \hat{\tilde x}_0(f, \theta) - \underset{f}{\text{argmax}} \,\, \tilde x_0(f, \theta)$.

\begin{figure}[t!]
    \centering
    \includegraphics[width=1\linewidth]{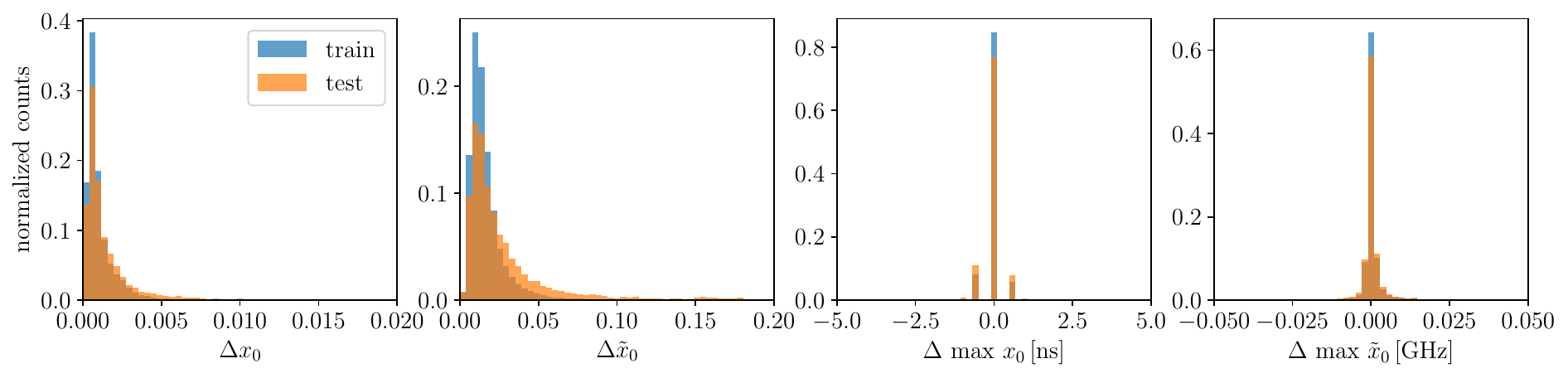}
    \caption{
    Performance of the $\theta$-Net.
    The distributions of the different evaluation metrics are depicted for training and test data.
    The distributions are almost the same, with only slightly heavier tails for the test data.
    }
    \label{fig:tNet_results}
\end{figure}

The distributions of these quantities are depicted in Figure \ref{fig:tNet_results}, for training and test data.
It is apparent that the model performs almost as well on the test data as on the training data, showing that there is no overfit.
The full evaluation metrics are shown in Table~\ref{tab:theta_eval}.
The 0.95-quantile of $0.006$ for $\Delta x_0$ shows that most signals are reproduced with an error of less than 1\%. 
Furthermore, the peaks of the signals are at the correct spot, or deviate by only one bin in the time domain in about 95\% of the cases, as evidenced by the results for $\Delta \text{max} \, x_0$.
Hence, the overall performance of the $\theta$-Net is excellent.

\subsection{Optimization experiment}

\begin{table}[t!]
\caption{
Evaluation metrics for the $\theta$-Net.
For positive quantities, the 0.95-quantile is given.
Otherwise, the 0.05- and 0.95-quantiles are given.
}
\label{tab:theta_eval}
\centering
\resizebox{0.7\textwidth}{!}{\begin{tabular}{lcccc}
\toprule
\textbf{$\theta$-Net} & $\Delta x_0$  & $\Delta \tilde x_0$ &  $\Delta \text{max } x_0 \, [\rm ns]$  &  $\Delta \text{max } \tilde x_0 \, [\rm GHz]$  \\
\midrule
train & 0.003 & 0.039 & [-0.500, 0.500] & [-0.008, 0.008] \\
test & 0.006 & 0.106 & [-0.500, 0.500] & [-0.023, 0.010] \\
\bottomrule
\end{tabular}}
\end{table}

\begin{figure}[t!]
    \centering
    \includegraphics[width=0.9\linewidth]{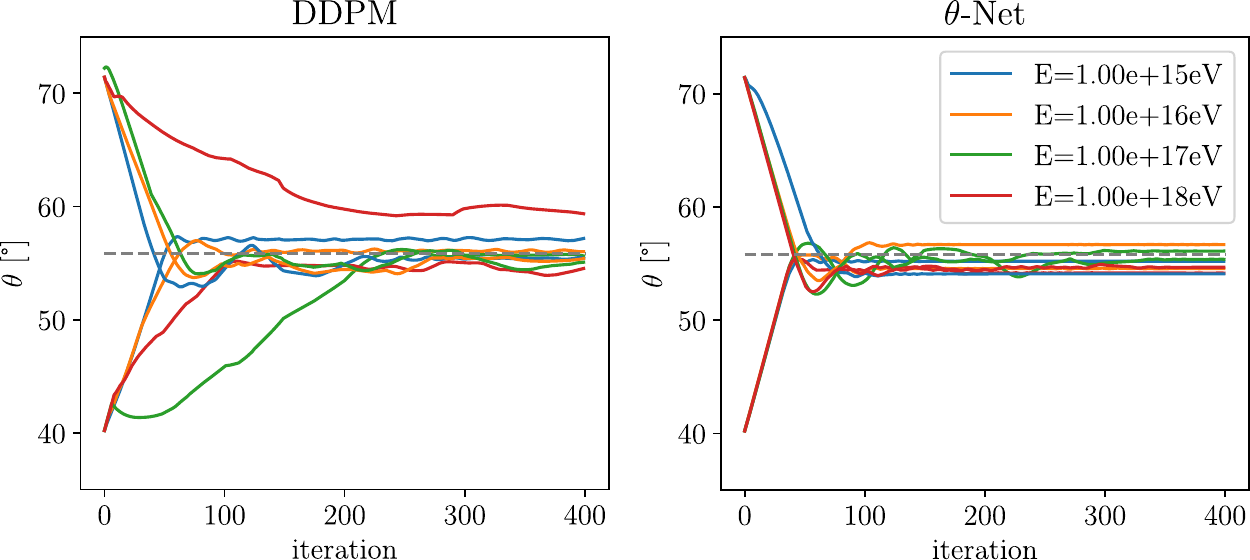}
    \caption{
    Optimizing the viewing angle to minimize the area under the normalized signals.
    The dashed line indicates the Cherenkov angle.
    \textbf{Left:}
    Using a vanilla DDPM.
    \textbf{Right:}
    Using MC data and modeling the angular dependence with the $\theta$-Net.
    }
    \label{fig:optimization}
\end{figure}

To evaluate the suitability of our model for usage as a constituent in a larger optimization pipeline, we also conducted a simple optimization experiment.
Here, the objective of the optimization is to minimize the area under the normalized signals, by optimizing $\theta$.
For this purpose, we employ the loss function
\begin{equation}
    A(\theta) = \frac{1}{N_{\rm bs}} \sum_{i=1}^{N_{\rm bs}} \int_{-\infty}^{\infty} |\hat x_{i0}(t, \theta)| \mathrm{d}t,
\end{equation}
where $N_{\rm bs}$ denotes the batch size.
The results are depicted in Figure \ref{fig:optimization}.
We conduct this experiment for an architecture with $\theta$-Net acting on the training data (i.e., using the MC simulation as generator), and an architecture with DDPM as generator, but without $\theta$-Net.

During the optimization, the viewing angle goes towards values close to the Cherenkov angle, as expected (compare Figure \ref{fig:theta_relation}).
When considering the pure DDPM variant without $\theta$-Net, the model can take considerably longer to converge at high energies.
Another advantage of the $\theta$-Net architecture is that the VRAM requirements during optimization are significantly lower, with about $0.8$GB, as opposed to around $3.6$GB when using the DDPM (with $N_{\rm bs} = 10$).
The execution speed is also much faster, with $0.06$s per iteration for the variant with $\theta$-Net vs $2.13$s per iteration for the variant with DDPM.
When combining the DDPM with $\theta$-Net, the VRAM advantage would persist since the computational graph would not need to be created for the DDPM, as explained in Section \ref{sec:model}.
The accuracy of the optimization results is limited by the resolution of the time grid.
The Adam optimizer with learning rate $0.01$ was used for this experiment, together with a learning rate scheduler halving the learning rate every 150 iterations.

\section{Conclusions and Outlook}
We have presented a differentiable surrogate model for the generation of radio pulses from in-ice neutrino interactions.
The model consists of three main components:
a generative model, a network to adjust the viewing angle of the signal, and a network to predict the signal amplitude.
This modularized model architecture allows for realistic and consistent sample generation across a wide range of energies and viewing angles.
In particular, it allows for the creation of related signals as measured by different antennas that stem from the same underlying event.
Moreover, the model remains differentiable even when using a non-differentiable generator, such as Monte Carlo simulations.

We have evaluated the different model components separately and demonstrated good model performance.
As of now, the best choice is to use existing Monte Carlo simulations for the generator.
While the diffusion model succeeds at reproducing the training data, more data would be required to enable the generation of truly novel samples.
Both the $\theta$-Net and the $a$-Net are highly accurate at their respective tasks.
Most importantly, the modularized model architecture has proven effective at enabling differentiability of the signals with regard to the parameters of interest in an accurate and hardware-efficient manner.

In the future, we aim to utilize the model developed in this work in a larger optimization pipeline to optimize the design of the radio detector in IceCube-Gen2.
We have already taken first steps in this direction for a simplified detector setup \citep{ravn2025optimization}.
Similar model architectures may also prove effective for enabling differentiability in unrelated applications, where the effect of the parameters of interest on the signal shape can be separated from the fundamental generative process.
A separate amplitude net may be useful for generative models in general, which deal with data that covers an extensive range of amplitudes.

\subsubsection*{Conflict of Interest statement}
There are no conflicts of interest.

\subsubsection*{Ethics statement}
There are no ethical concerns.

\subsubsection*{Funding statement}
The work is financially supported by the Swedish Research Council (VR) via the project \emph{Physics-informed machine learning} (registration number: 2021-04321).
CG is supported by the European Union (ERC, NuRadioOpt, 101116890).
This work used resources provided by the National Academic Infrastructure for Supercomputing in Sweden (NAISS), partially funded by the Swedish Research Council through grant agreement no. 2022-06725.

\newpage
\bibliographystyle{unsrtnat}
\bibliography{ice}

\newpage
\appendix

\section{Ablation study for the $\theta$-Net} \label{app:tnet_ablation}

In this appendix, we conduct an ablation study to demonstrate the impact that the different components of the $\theta$-Net depicted in Figure \ref{fig:tNet} have.
The evaluation metrics introduced in Section \ref{sec:tnet_eval} are given in Table \ref{tab:theta_ablation} for the $\theta$-Net with different components missing.

The preprocessing without any further fine-tuning performs clearly worse than the full $\theta$-Net.
When omitting the preprocessing, the resulting model also performs poorly.
This demonstrates that the U-Net benefits from being given a simpler task to learn.
A likely reason for this is that long-range dependencies, which are relevant when transforming signals from large to small angles, can be difficult to learn for convolutional architectures.
Omitting the CNN for extracting additional features only slightly degrades the performance.
Its inclusion is therefore not essential, but still beneficial.

\begin{table}[t]
\caption{
Ablation study for the $\theta$-Net.
For positive evaluation metrics, the 0.95-quantile is given.
Otherwise, the 0.05- and 0.95-quantiles are given.
Bold font highlights best performance.
}
\label{tab:theta_ablation}
\centering
\resizebox{0.8\textwidth}{!}{\begin{tabular}{lcccc}
\toprule
 & $\Delta x_0$  & $\Delta \tilde x_0$ &  $\Delta \text{max } x_0 \, [\rm ns]$  &  $\Delta \text{max } \tilde x_0 \, [\rm GHz]$  \\
\midrule
only preprocessing & 0.020 & 0.237 & [-1.500, 2.000] & [-0.045, 0.113] \\
no preprocessing & 0.028 & 0.176 & [-1.500, 1.000] & [-0.039, 0.020] \\
no CNN & 0.010 & 0.107 & [\textbf{-0.500}, 1.000] & [\textbf{-0.016}, 0.012] \\
full $\theta$-Net & \textbf{0.006} & \textbf{0.106} & [\textbf{-0.500}, \textbf{0.500}] & [-0.023, \textbf{0.010}] \\
\bottomrule
\end{tabular}}
\end{table}

\section{Hyperparameter selection} \label{app:hypeparameters}

The final choices for the details of the model architecture stated in Section \ref{sec:architecture} were guided by the requirements of computational efficiency and accuracy.
In general, the number of network layers and their respective numbers of channels and widths were chosen as small as possible without negatively impacting the validation accuracy.
For the model components involving a U-Net in particular, the number of levels, the depth of the ResNet blocks, as well as the number of channels were tuned.
Omitting the self-attention layer, which is sometimes applied after the final ResNet block of each level in U-Net architectures led to significant improvements in both the network size and the model performance.
The presented model outperformed variants of the architecture with both roughly half as many and twice as many parameters.

The training hyperparameters were chosen to ensure effective and efficient training.
We manually tuned the learning rates to train the different model components fast, without negatively affecting their validation accuracy.
Increasing the learning rates by an order of magnitude over the chosen values typically led to decreased performance.
The number of iterations was chosen to stop the training after convergence.

\section{Background on diffusion models} \label{app:ddpm}

Denoising diffusion probabilistic models (DDPMs) \citep{ho2020denoising} have gained a large amount of popularity since they were first introduced.
They have been successfully used in a large variety of applications \citep{yang2023diffusion}, most prominently in computer vision \citep{croitoru2023diffusion}.
A major advantage of DDPMs over other generative models, such as generative adversarial models (GANs), is that the training is more robust due to their likelihood-based objective function.
They are also less prone to suffering from issues such as mode collapse.

When training DDPMs, two processes are involved.
In the forward or diffusion process, noise is added step-by-step to samples $x_0$ from the training data:
\begin{align}
q(x_k \mid x_{k-1}) &= \mathcal{N}(x_k; \sqrt{1 - \beta_k} \, x_{k-1}, \beta_k \text{I}), \\
q(x_k \mid x_0) &= \mathcal{N}(x_k; \sqrt{\bar{\alpha}_k} \, x_0, (1 - \bar{\alpha}_k) \text{I}),
\end{align}
where the $\beta_k$ define the noise schedule, $x_k$ denotes the sample after adding noise $k$ times, and where $\alpha_k = 1 - \beta_k$, $\bar{\alpha}_k = \prod_{s=1}^k \alpha_s$.
In the reverse or denoising process, the diffusion model reverts this process and removes the noise step-by-step:
\begin{equation}
p_\theta(x_{k-1} \mid x_k) = \mathcal{N}(x_{k-1}; \mu_\theta(x_k, k), \beta_k \text{I}),
\end{equation}
where $x_k = \sqrt{\bar{\alpha}_k} \, x_0 + \sqrt{1 - \bar{\alpha}_k}$ and $\epsilon \sim \mathcal{N}(0, \mathbf{I})$, and where $\mu_\theta$ denotes the predicted mean.
In practice, however, it is more effective to let the diffusion model predict the noise directly. The model is then trained with the loss function
\begin{equation}
\mathcal{L} = \mathbb{E}_{x_0, \epsilon, k} \left[ 
\left\| \epsilon - \epsilon_\theta\left( \sqrt{\bar{\alpha}_k} \, x_0 + \sqrt{1 - \bar{\alpha}_k} \, \epsilon, \, k \right) \right\|^2 \right],
\end{equation}
where $\epsilon_\theta$ denotes the noise prediction of the denoiser, which constitutes the main component of the DDPM that needs to be learned.
Once fully trained, the DDPM can generate samples from random Gaussian noise.

\subsection{Generated samples}
\begin{figure}[t]
    \centering
    \includegraphics[width=\linewidth]{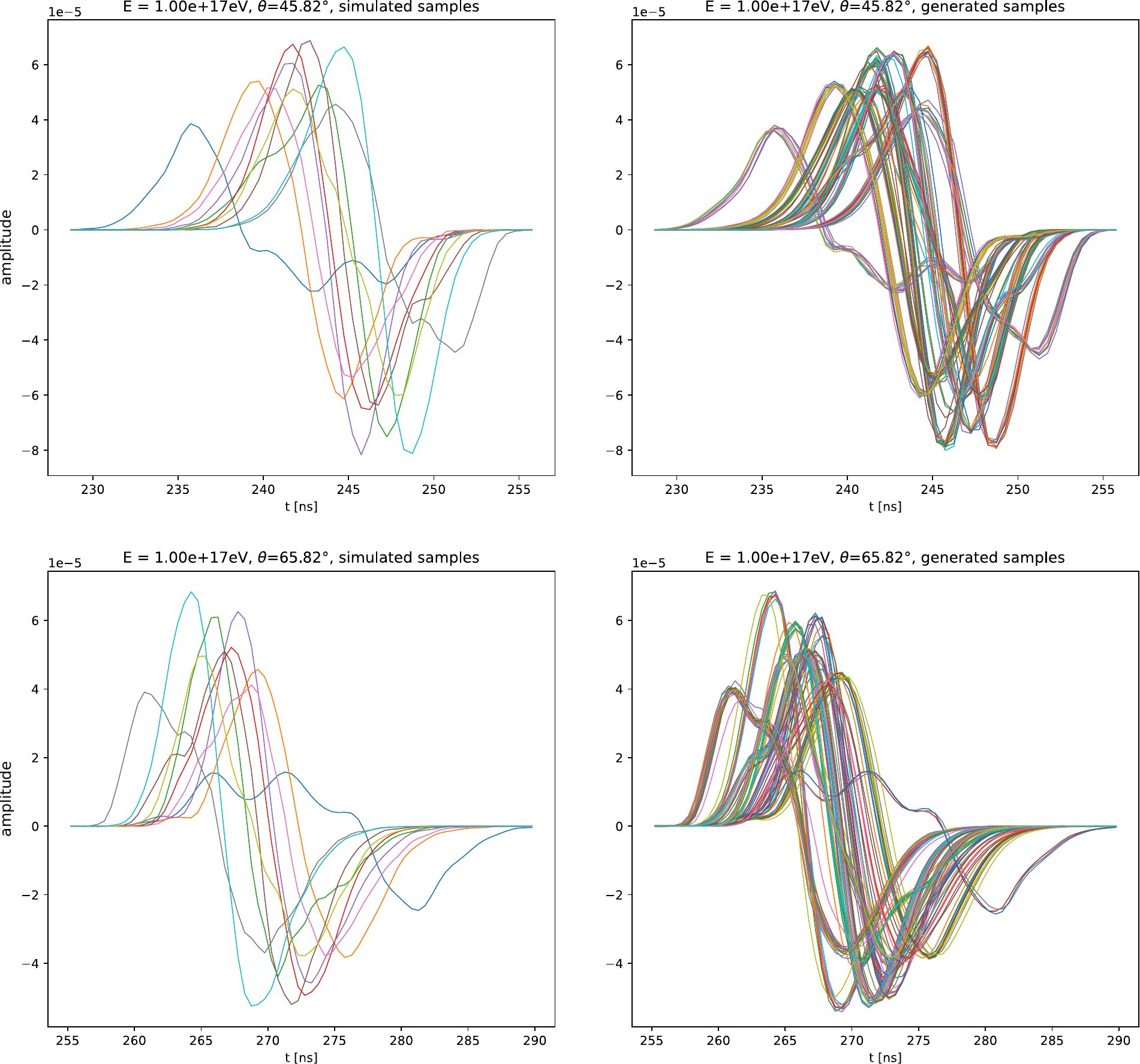}
    \caption{
    Radio signals at different energies $E$ and viewing angles $\theta$.
    \textbf{Left:} Samples from the Monte Carlo simulation.
    \textbf{Right:} Samples generated by the diffusion model.
    }
    \label{fig:DDPM_samples}
\end{figure}

In Figure \ref{fig:DDPM_samples}, generated samples from the diffusion model at fixed $E$ and $\theta$ values are compared to corresponding samples from the training data.
It is apparent that the DDPM accurately reproduces the training data.
This shows that the DDPM is well-suited for modeling this type of data.
However, while the generated samples exhibit a certain variation around the training samples, the DDPM does not yet produce truly novel signals.
This is presumably due to the limited size of the training data, which contains only 410 fundamentally different samples.
There are efforts underway to generate a larger dataset, which may enable the DDPM to go beyond merely reproducing the training data in the future.

\begin{figure}[t]
    \centering
    \includegraphics[width=\linewidth]{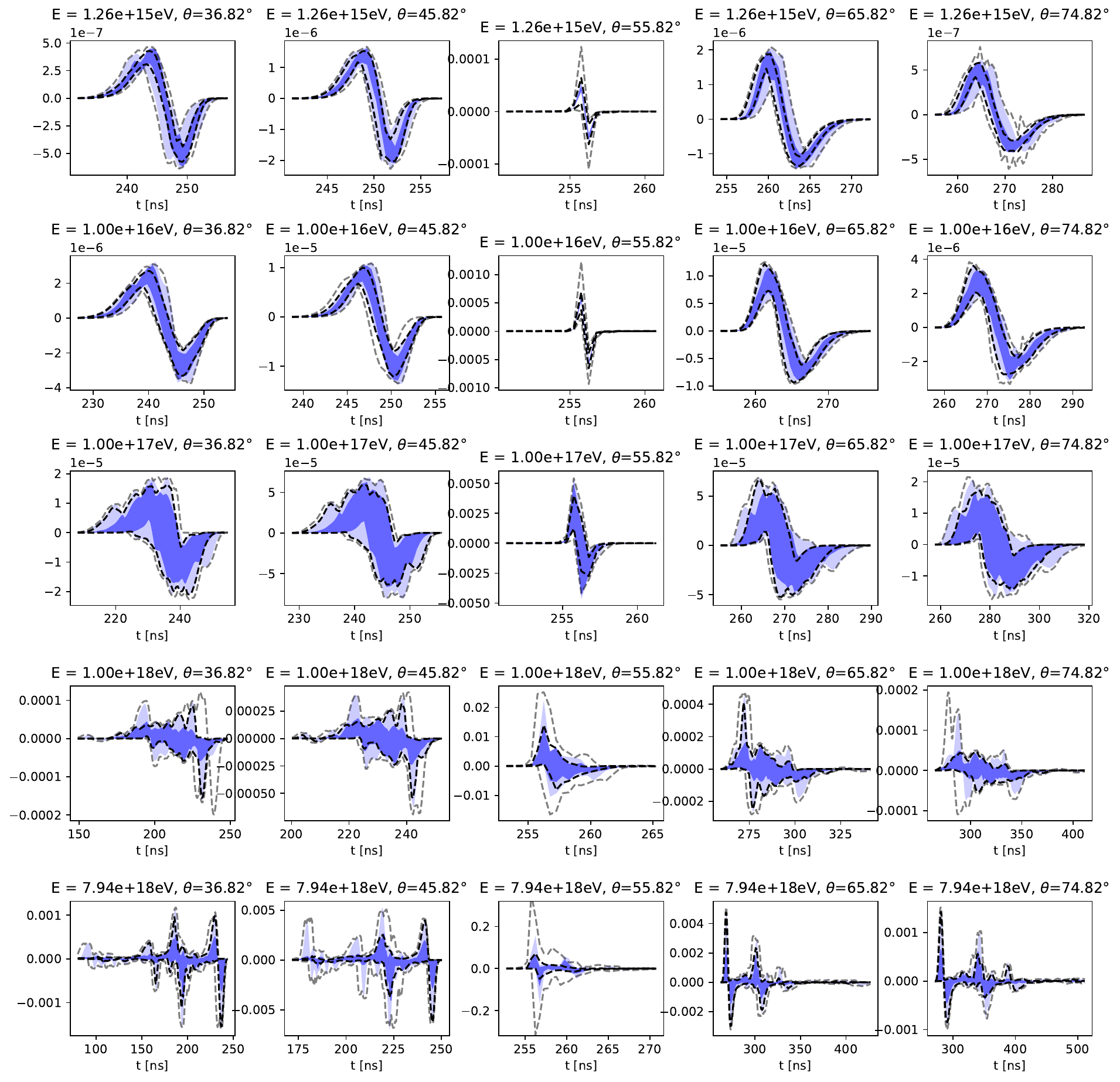}
    \caption{
    The distribution of real and generated signals, conditional on the energy $E$ and angle $\theta$ (within $\pm0.025 \Delta_{E/\theta}$ of the corresponding values in logspace).
    The values on the y-axis correspond to the signal amplitude.
    The shaded blue area corresponds to the real distribution, with the light blue area corresponding to the total range of signal values at that point and the dark blue area to the range from the 0.1-quantile to the 0.9-quantile.
    The dashed lines depict the corresponding quantiles for the generated data.
    }
    \label{fig:DDPM_contours}
\end{figure}

In Figure \ref{fig:DDPM_contours}, the distribution of the existing data and the generated data is depicted around different energies $E$ and viewing angles $\theta$.
It is apparent that, on the whole, the distributions agree well, indicating that the DDPM reproduces the entire training data and does not suffer from mode collapse.

\end{document}